\documentclass[final,3p,times]{elsarticle}

\usepackage{graphicx}

\usepackage{amssymb}
\usepackage{amsthm}

\usepackage{hyperref}
\usepackage{verbatim}

 \biboptions{square,sort&compress,comma}

\journal{Computer Physics Communications}

\usepackage{mathrsfs}
\usepackage{amssymb}
\usepackage{amsmath}
\usepackage{bm}
\usepackage{graphicx}
\usepackage{color}

\begin{document}

\begin{frontmatter}



\title{PDRF: A General Dispersion Relation Solver for Magnetized Multi-Fluid Plasma}

\date{\today}


\author{Hua-sheng XIE}
\ead{huashengxie@gmail.com}

\address{Institute for Fusion Theory and Simulation, Zhejiang
University, Hangzhou, 310027, PRC}

\begin{abstract}
A general dispersion-relation solver that
numerically evaluates the full propagation properties of all the waves in fluid plasmas is presented. The effects of anisotropic pressure, external magnetic fields and beams, relativistic dynamics, as well as local plasma inhomogeneity are included. [Computer Physics Communications, (2013); doi: 10.1016/j.cpc.2013.10.012; code: http://cpc.cs.qub.ac.uk/summaries/AERF\_v1\_0.html]\\
\bigskip \\
\textbf{Program summary}\\
\\
\emph{Title of program:} PDRF \\
\emph{Catalogue identifier: AERF\_v1\_0}   \\
\emph{Program summary URL: http://cpc.cs.qub.ac.uk/summaries/AERF\_v1\_0.html}  \\
\emph{Program obtainable from:} CPC Program Library, Queen
University of Belfast, N.
 Ireland\\
\emph{Computer for which the program is designed and others on
which it
has been tested: Computers:} Any computer running \emph{MATLAB} 7. Tested on DELL OptiPlex 380.\\
\emph{Installations:} Institute for Fusion Theory and Simulation,
Zhejiang University, Hangzhou, PRC\\
\emph{Operating systems under which the program has been tested:} Windows XP Pro\\
\emph{Programming language used:} \emph{MATLAB} 7
\\
\emph{Memory required to execute with typical data:} 10 M\\
\emph{No. of lines in distributed program, including test data,
etc.:} 340 \\
\emph{No. of bytes in distributed program, including test data,
etc.:} 12 000 \\
\emph{Distribution format:} .tar.gz \\
\emph{Nature of physical problem:} This dispersion relation solver
calculates all the solutions and gives corresponding polarizations
for magnetized fluid plasma with arbitrary number of components and
arbitrary orient wave vector, with and without anisotropic pressure,
relativistic,
beam and local inhomogeneity effects. \\
\emph{Method of solution:} Solving as matrix eigenvalue problem.\\
\emph{Restrictions on the complexity of the problem:} No kinetic and nonlinear effects.\\
\emph{Typical running time:} About 1 second on a Intel Pentium
2.60 GHz PC. \\
\emph{Unusual features of the program:} Giving all the solutions and
polarizations.

\end{abstract}


\end{frontmatter}


\section{Introduction}\label{sec:intro}
Since only a few relatively simple dispersion relations in plasma
physics are analytical tractable, it is of practical interest to
develop general numerical schemes for obtaining and solving the
dispersion relations of given plasma systems. Traditionally, one
obtains the dispersion relations from the determinant of the
corresponding dielectric tensor for the eigenvalues of the wave
solutions, such as that of the kinetic code WHAMP (Waves in
Homogeneous Anisotropic Multicomponent Magnetized Plasma) by
Ronnmark\cite{Ronnmark1982,Ronnmark1983} and the {\it Mathematica}
fluid code for magnetized parallel beam plasmas by
Bret\cite{Bret2007}. It is however difficult to generalize such
treatments to include arbitrary number of fluid species with good
convergence or to obtain all the solutions of a given system.

In this work, we use a full-matrix approach to transform the task of
obtaining the dispersion relations to an equivalent matrix
eigenvalue problem by introducing a general dispersion relation
solver PDRF (Plasma Dispersion Relation - Fluid Version) for
magnetized multi-fluid plasmas including anisotropic, relativistic,
beam and weak inhomogeneity effects. Our solver should be useful for
investigating wave propagation properties in astrophysical, space,
laser, and laboratory plasmas.

\section{Theory and Method}\label{sec:meth}
We consider a multi-fluid plasma in an external magnetic field $\bm
B_0=(0,0,B_0)$. The flow velocity of the fluid component $j$ is $\bm
v_{j0}=(v_{j0x},v_{j0y},v_{j0z})$. The species densities are allowed
to be locally inhomogeneous, with $\nabla
n_{j0}/n_{j0}=(\epsilon_{njx},\epsilon_{njy},0)=$ constant. For
simplicity, temperature gradient effects are ignored, and the wave
vector is assumed to be
$\bm{k}=(k_x,0,k_z)=(k\sin\theta,0,k\cos\theta)$. We do not need to
obtain the 3 by 3 dispersion relation matrix for $\bm
E=(E_x,E_y,E_z)$, as done in many existing analytical or numerical
treatments, such as in Stix\cite{Stix1992},
Ronnmark\cite{Ronnmark1982,Ronnmark1983} and Bret\cite{Bret2007}.
Instead, we use the original full dispersion relation matrix and
treat it as a matrix eigenvalue problem, instead of directly solving
for its determinant.

\subsection{Governing equations}\label{sec:fpdr}
We start with the fluid equations
\begin{subequations} \label{eq:fpeq}
\begin{eqnarray}
  & \partial_t n_j = -\nabla\cdot(n_j\bm v_j),\\
  & \partial_t \bm u_j = -\bm v_j\cdot \nabla\bm u_j+\frac{q_j}{m_j}(\bm E+\bm v_j\times \bm B)-\frac{\nabla\bm P_j}{\rho_j}-\sum_i(\bm u_i-\bm u_j)\nu_{ij},\\
  & \partial_t \bm E = c^2\nabla\times\bm B - \bm J/\epsilon_0,\\
  & \partial_t \bm B = -\nabla\times\bm E,
\end{eqnarray}
\end{subequations}
where $\bm u_j=\gamma_j \bm v_j$, and
\begin{subequations} \label{eq:fpeq2}
\begin{eqnarray}
  & \bm J = \sum_jq_jn_j\bm v_j, \\
  & d_t(P_{\parallel j}\rho_j^{-\gamma_{\parallel j}}) = 0, \\
  & d_t(P_{\perp j}\rho_j^{-\gamma_{\perp j}}) = 0,
\end{eqnarray}
\end{subequations}
where $\rho_{j}\equiv m_jn_{j}$, $c^2=1/\mu_0\epsilon_0$,
$\gamma_j=(1-v_j^2/c^2)^{-1/2}$, and $\gamma_{\parallel j}$ and
$\gamma_{\perp j}$ are the parallel and perpendicular adiabatic
coefficients, respectively. Furthermore,
$P_{\parallel,\perp}=nT_{\parallel,\perp}$, $\bm
P=P_{\parallel}\hat{\bm b}\hat{\bm b}+P_{\perp}(\bm I-\hat{\bm
b}\hat{\bm b})$ and $\hat{\bm b}=\bm B/B$. Note that our
anisotropy model differs from that of CGL\cite{Chew1956},
but can be reduced to that of Bret and Deutsch\cite{Bret2006} by
setting $\gamma_{\parallel j}=\gamma_{\perp j}=\gamma_{T j}$. By
further setting $T_{\perp j}=T_{\parallel j}$, we can recover the
isotropic pressure case.

After linearizing, (\ref{eq:fpeq2}) becomes
\begin{subequations} \label{eq:fpeq2lin}
\begin{eqnarray}
  & \bm J = \sum_jq_j(n_{j0}\bm v_{j1}+n_{j1}\bm v_{j0}), \\
  & P_{\parallel,\perp j1} = c^2_{\parallel,\perp j}n_{j1},
\end{eqnarray}
\end{subequations}
where $c^2_{\parallel,\perp j}=\gamma_{\parallel,\perp
j}P_{\parallel,\perp j0}/\rho_{j0}$ and $\bm P_{j0}=n_{j0}\bm
T_{j0}$.

We note that
\begin{equation} \label{eq:gradP}
   \nabla\cdot\bm P_{j1} = (ik_x,0,ik_z)\cdot\left[\begin{array}{ccc}
    P_{\perp j1} & 0 & \Delta_j B_{x1}\\
    0 & P_{\perp j1} & \Delta_j B_{y1}\\
    \Delta_j B_{x1} & \Delta_j B_{y1} & P_{\parallel j1}
    \end{array}
 \right],
\end{equation}
where $\Delta_j\equiv(P_{\parallel j0}-P_{\perp j0})/B_0$ and
$\beta_{\parallel,\perp j}=2\mu_0P_{\parallel,\perp j}/B_0^2$. The
off-diagonal terms coming from the tensor rotation (see similar
expressions in \ref{sec:pdrfu}) from $\hat{\bm b}_0$ to $\hat{\bm
b}$ are related to energy exchange and are important for the
anisotropic instabilities. An incorrect treatment or ignoring these
off-diagonal terms can cause loss of the firehose and other unstable
anisotropic modes.

\subsection{Full-matrix treatment}
The linearized version of (\ref{eq:fpeq}) with
$f=f_0+f_1e^{i\bm{k\cdot r}-i\omega t}$, $f_1\ll f_0$ is equivalent to a matrix
eigenvalue problem
\begin{equation}\label{eq:eig}
    \lambda\bm A\bm
X=\bm M\bm X,
\end{equation}
with $\lambda=-i\omega$ the eigenvalue and $\bm X$ the corresponding
eigenvector, which also gives the polarization of each
normal/eigen mode solution. Similar treatments can be found in
\cite{Goedbloed2004} for the MHD equations and \cite{Hakim2008} for
the ten-moment equations.

Accordingly, we have $\bm X=(n_{j1},v_{j1x},v_{j1y},v_{j1z},E_{1x},E_{1y},E_{1z},B_{1x},B_{1y},B_{1z})^T$,
$\bm u_{j1}=\gamma_{j0}[\bm v_{j1}+\gamma_{j0}^2(\bm
v_{j0}\cdot\bm v_{j1})\bm v_{j0}/c^2]=\{a_{jpq}\}\cdot\bm v_{j1}$
($p,q=x,y,z$), $\gamma_{j0}=(1-v_{j0}^2/c^2)^{-1/2}$, and $\bm
A$ is given by
\begin{equation}\label{eq:fpdrA}
\left[\begin{array}{cccccccccc}
    \{1 & 0 & 0 & 0  & 0 & 0 & 0 & 0 & 0 & 0\\
    0 & a_{jxx} & a_{jxy} & a_{jxz}  & 0 & 0 & 0 & 0 & 0 & 0\\
    0 & a_{jyx} & a_{jyy} & a_{jyz}  & 0 & 0 & 0 & 0 & 0 & 0\\
    0 & a_{jzx} & a_{jzy} & a_{jzz}\}  & 0 & 0 & 0 & 0 & 0 & 0\\
    0 & 0 & 0 & 0  & 1 & 0 & 0 & 0 & 0 & 0\\
    0 & 0 & 0 & 0  & 0 & 1 & 0 & 0 & 0 & 0\\
    0 & 0 & 0 & 0  & 0 & 0 & 1 & 0 & 0 & 0\\
    0 & 0 & 0 & 0  & 0 & 0 & 0 & 1 & 0 & 0\\
    0 & 0 & 0 & 0  & 0 & 0 & 0 & 0 & 1 & 0\\
    0 & 0 & 0 & 0  & 0 & 0 & 0 & 0 & 0 & 1
    \end{array}
 \right].
\end{equation}

For simplicity, the relativistic effects in the friction terms
$\nu_{ij}$ are ignored. The matrix $\bm M$ is then given by
($\nu_{ij}$ terms when $i\neq j$ are not given explicitly here)
\begin{equation}\label{eq:fpdrM}
\left[\begin{array}{cccccccccc}
    \{-i\bm k\cdot\bm v_{j0} & -ik_xn_{j0}-\epsilon_{njx}n_{j0} & -\epsilon_{njy}n_{j0} & -ik_zn_{j0} & 0 & 0 & 0 & 0 & 0 & 0\\
    \frac{-ik_xc_{\perp j}^2}{\rho_{j0}} & b_{jxx} & b_{jxy}+\omega_{cj} & b_{jxz}  & \frac{q_j}{m_j} & 0 & 0 & -\frac{ik_z\Delta_j}{m_jn_{j0}} & -\frac{q_jv_{j0z}}{m_j} & \frac{q_jv_{j0y}}{m_j}\\
    0 & b_{jyx}-\omega_{cj} & b_{jyy} & b_{jyz}  & 0 & \frac{q_j}{m_j} & 0 & \frac{q_jv_{j0z}}{m_j} & -\frac{ik_z\Delta_j}{m_jn_{j0}} & -\frac{q_jv_{j0x}}{m_j}\\
    \frac{-ik_zc_{\parallel j}^2}{\rho_{j0}} & b_{jzx} & b_{jzy} & b_{jzz}\}  & 0 & 0 & \frac{q_j}{m_j} & -\frac{q_jv_{j0y}}{m_j}-\frac{ik_x\Delta_j}{m_jn_{j0}} & \frac{q_jv_{j0x}}{m_j} & 0\\
    -\frac{q_jv_{j0x}}{\epsilon_0} & -\frac{q_jn_{j0}}{\epsilon_0} & 0 & 0  & 0 & 0 & 0 & 0 & -ik_zc^2 & 0\\
    -\frac{q_jv_{j0y}}{\epsilon_0} & 0 & -\frac{q_jn_{j0}}{\epsilon_0} & 0  & 0 & 0 & 0 & ik_zc^2 & 0 & -ik_xc^2\\
    -\frac{q_jv_{j0z}}{\epsilon_0} & 0 & 0 & -\frac{q_jn_{j0}}{\epsilon_0}  & 0 & 0 & 0 & 0 & ik_xc^2 & 0\\
    0 & 0 & 0 & 0  & 0 & ik_z & 0 & 0 & 0 & 0\\
    0 & 0 & 0 & 0  & -ik_z & 0 & ik_x & 0 & 0 & 0\\
    0 & 0 & 0 & 0  & 0 & -ik_x & 0 & 0 & 0 & 0
    \end{array}
 \right],
\end{equation}
where $\omega_{cj}=q_jB_0/m_j$, $q_e=-e$,
$\omega_{pj}^2=n_{j0}q_j^2/\epsilon_0m_j$, and
$\{b_{jpq}\}=\nu_{jj}-i(\bm k\cdot\bm v_{j0})\cdot\{a_{jpq}\}$. Furthermore,
\begin{eqnarray}\label{eq:ajxyz}
\{a_{jpq}\}\equiv\left[\begin{array}{ccc}
    a_{jxx} & a_{jxy} & a_{jxz}\\
    a_{jyx} & a_{jyy} & a_{jyz}\\
    a_{jzx} & a_{jzy} & a_{jzz}
    \end{array}
 \right]=\left[\begin{array}{ccc}
    \gamma_{j0}+\gamma_{j0}^3v_{j0x}^2/c^2 & \gamma_{j0}^3v_{j0x}v_{j0y}/c^2 & \gamma_{j0}^3v_{j0x}v_{j0z}/c^2\\
    \gamma_{j0}^3v_{j0x}v_{j0y}/c^2 & \gamma_{j0}+\gamma_{j0}^3v_{j0y}^2/c^2 & \gamma_{j0}^3v_{j0y}v_{j0z}/c^2\\
    \gamma_{j0}^3v_{j0x}v_{j0z}/c^2 & \gamma_{j0}^3v_{j0z}v_{j0y}/c^2 & \gamma_{j0}+\gamma_{j0}^3v_{j0z}^2/c^2
    \end{array}
 \right].
\end{eqnarray}

For a plasma containing $s$ species of particles, the dimensions of $\bm A$ and
$\bm M$ are $(4s+6)\times(4s+6)$. We can get all the linear harmonic wave
solutions of the system without any convergence difficulty using a
standard matrix eigenvalue solver, e.g., LAPACK or the function {\it
eig()} in {\it MATLAB}. Here, a {\it MATLAB} code PDRF for solving
the above eigenvalue problem is provided. By setting $\gamma_j$ to
$1$, i.e., $\bm A=\bm I$ and $\{a_{jpq}\}=\bm I$, PDRF reduces to
the non-relativistic case.

\subsection{Validity} 
For the perturbation analysis, we have assumed
$Q_0=\sum_jq_jn_{j0}\sim0$, $\bm J_0=\sum_jq_jn_{j0}\bm
v_{j0}\sim 0$, $\bm v_{j0}\cdot\nabla n_{j0}\sim0$ and $\bm
F_{j0}=q_j\bm E_0+q_j(\bm v_{j0}\times\bm B_0)-\nabla\bm
P_{j0}/n_{j0}-m_j\sum_i(\bm v_{i0}-\bm v_{j0})\nu_{ij}\sim0$.
In practice, these often-used assumptions may not always be valid. For
example, a finite external current $\bm J_0$ would be associated with an external $\bm B_{0\bm J}$ field. To determined the latter, we will need the boundary conditions
or other information. That could also mean that the system is no longer homogenous. For example, a local current can generate an inhomogeneous magnetic field around it. Such finite lowest-order quantities cannot be removed by simple coordinate
transformations. Similar difficulties can appear for non-parallel external beams.
On the other hand, finite $\bm v_{\perp0}$ is allowed if there are external electric fields, local inhomogeneities, gravitational force, etc. in the perpendicular direction.
Moreover, when global gradient or inhomogeneity effects are included, the system will not be near uniform and a full treatment including the space dependent lowest order terms as well as the appropriate boundary conditions are required. However, here we are interested only in the local perturbations, so that the harmonic analysis, namely an $e^{i\bm{k\cdot r}-i\omega t}$ dependence for the first order (linear) quantities, can still be used \cite{Stix1992}). Possible breakdowns of the assumptions in PDRF are summarized in Table
\ref{tab:break}.

\begin{table}\begin{center}
\caption{\label{tab:break} Validity of the linearization.}
\begin{tabular}{cccccccc}\\\hline\hline
 - & 1 & 2  & 3  & 4  & 5  & 6 & 7 \\\hline
 Case $\neq 0$ & $\sum_jq_jn_{j0}$ & $\bm \sum_jq_jn_{j0}\bm
v_{j0}$ & $\bm E_0$ & $\nabla\bm P_{j0}$ & $\bm v_{j0}\times\bm B_0
$ & $\sum_i(\bm v_{i0}-\bm v_{j0})\nu_{ij}$ & $\bm v_{j0}\cdot\nabla
n_{j0}$ \\\hline\hline
\end{tabular}
\end{center}\end{table}

\section{Benchmarks}\label{sec:bech}
\subsection{Cold plasma} 
\begin{figure}
\begin{center}
  \includegraphics[width=16cm]{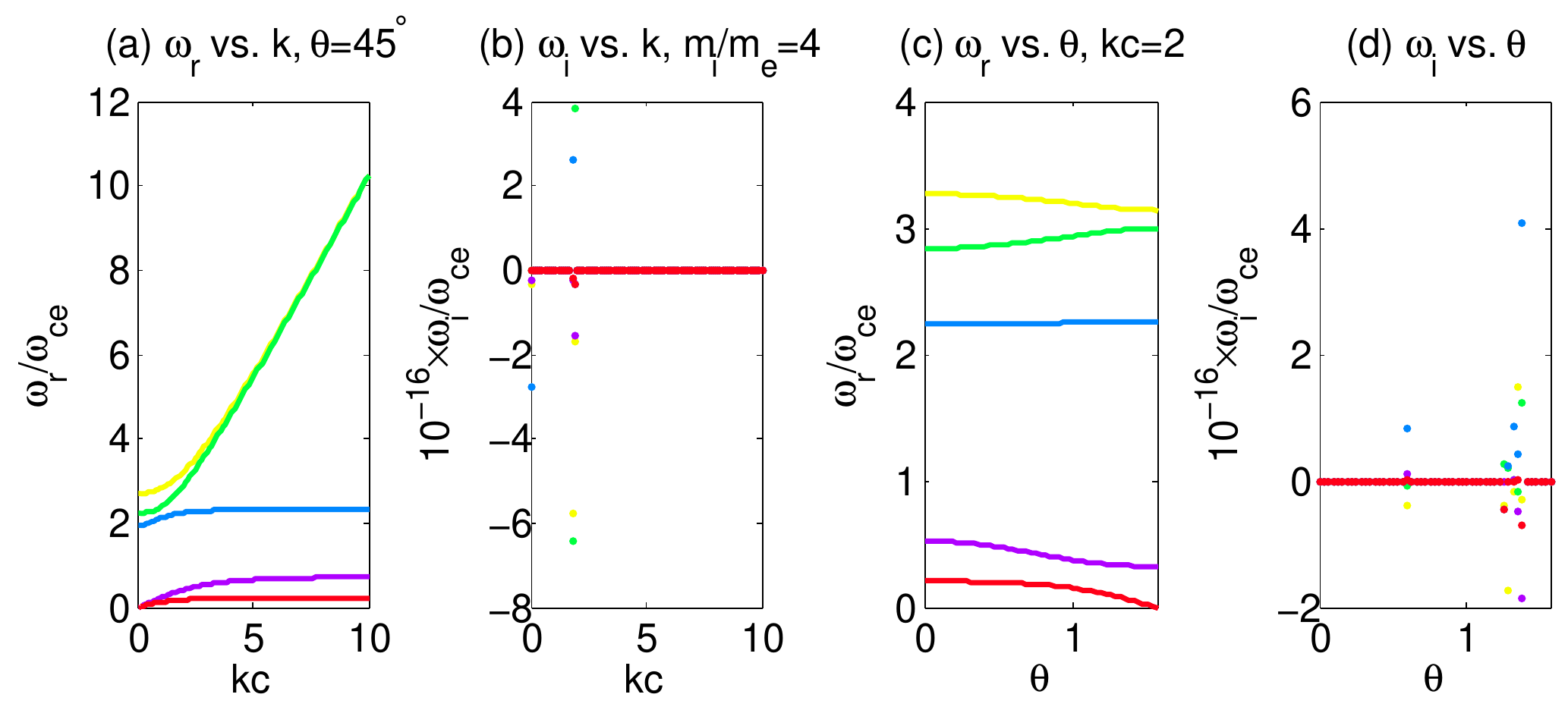}\\
  \caption{The dispersion curves $\omega_{r,i}$ versus $k$ and $\theta$ for $m_i/m_e=4$ and $\omega_{pe}=2\omega_{ce}$.}\label{fig:coldplasma}
\end{center}
\end{figure}

The numerical solutions $\omega^M$ of (\ref{eq:eig}) and that
$\omega^S$ of Swanson's fifth order polynomial
method\cite{Swanson2003} for a cold two-component plasma in the
absence of an external beam are shown in Table
\ref{tab:clodswanson}, for $kc=0.1$, $\theta=\pi/3$, $m_i/m_e=1836$
and $\omega_{pe}=10\omega_{ce}$ (Note: hereafter,
$\omega_{cj}=|\omega_{cj}|$). We see that the two results are fully
consistent. The small difference ($<10^{-15}$) can be attributed to
numerical errors. The dispersion curves for $\omega_{r,i}$ versus
$k$ and $\theta$ are given in Fig.\ \ref{fig:coldplasma}, for
$m_i/m_e=4$ and $\omega_{pe}=2\omega_{ce}$.

\begin{table}
\begin{center}
\caption{\label{tab:clodswanson} Comparison of the cold plasma solutions
using the matrix method and Swanson's polynomial method\cite{Swanson2003}.}
\begin{tabular}{cccccccc} \\\hline\hline
  $\omega^M$ & $\pm$10.5152 & $\pm$10.0031 & $\pm$9.5158 & $\pm$(2.4020E-4-i8E-19) &  $\pm$(1.1330E-4-i1E-16) & $\pm$0 & $\pm$0 \\\hline
  $\omega^S$ & $\pm$10.5152 & $\pm$10.0031 & $\pm$9.5158 &  $\pm$2.4020E-4 &  $\pm$1.1330E-4 &  - & - \\\hline\hline
\end{tabular}
\end{center}
\end{table}

\subsection{The firehose and mirror instabilities}
Two well-known low-frequency hydromagnetic anisotropic instabilities
are the firehose and mirror instabilities. The dispersion relation
for the firehose (or, anisotropic shear Alfv\'en) mode is
$(\frac{\omega}{k_zv_A})^2\simeq 1+\sum_j\frac{\beta_{\perp
j}-\beta_{\parallel j}}{2}$. For $k_z\ll k_\perp$, using the reduced
expression from the bi-Maxwellian dispersion relation, we
can obtain the mirror instability threshold $1+\sum_j\beta_{\perp
j}\big(1-\frac{\beta_{\perp j}}{\beta_{\parallel j}}\big)<0$
\cite{Hasegawa1975}. A correction (replacing $\Delta_j$ by
$2\Delta_j$) in the matrix element $M_{v_{j1z},B_{x1}}$ is required
to give the same mirror instability threshold of the kinetic
bi-Maxwellian plasma prediction since there is an extra factor $2$
in the bi-Maxwellian expression\cite{Southwood1993}, i.e.,
\begin{equation}\label{eq:mm2}
    \delta p_{\perp}=2p_{\perp}(1-\frac{p_{\perp}}{p_{\parallel}})\delta
    B_{\perp},
\end{equation}
when compared to (\ref{eq:gradP}). A similar difference for the mirror
instability solutions of the single-fluid MHD and the bi-Maxwellian
kinetic equations also exists \cite{Hau1993}.

\begin{figure}
\begin{center}
  \includegraphics[width=12cm]{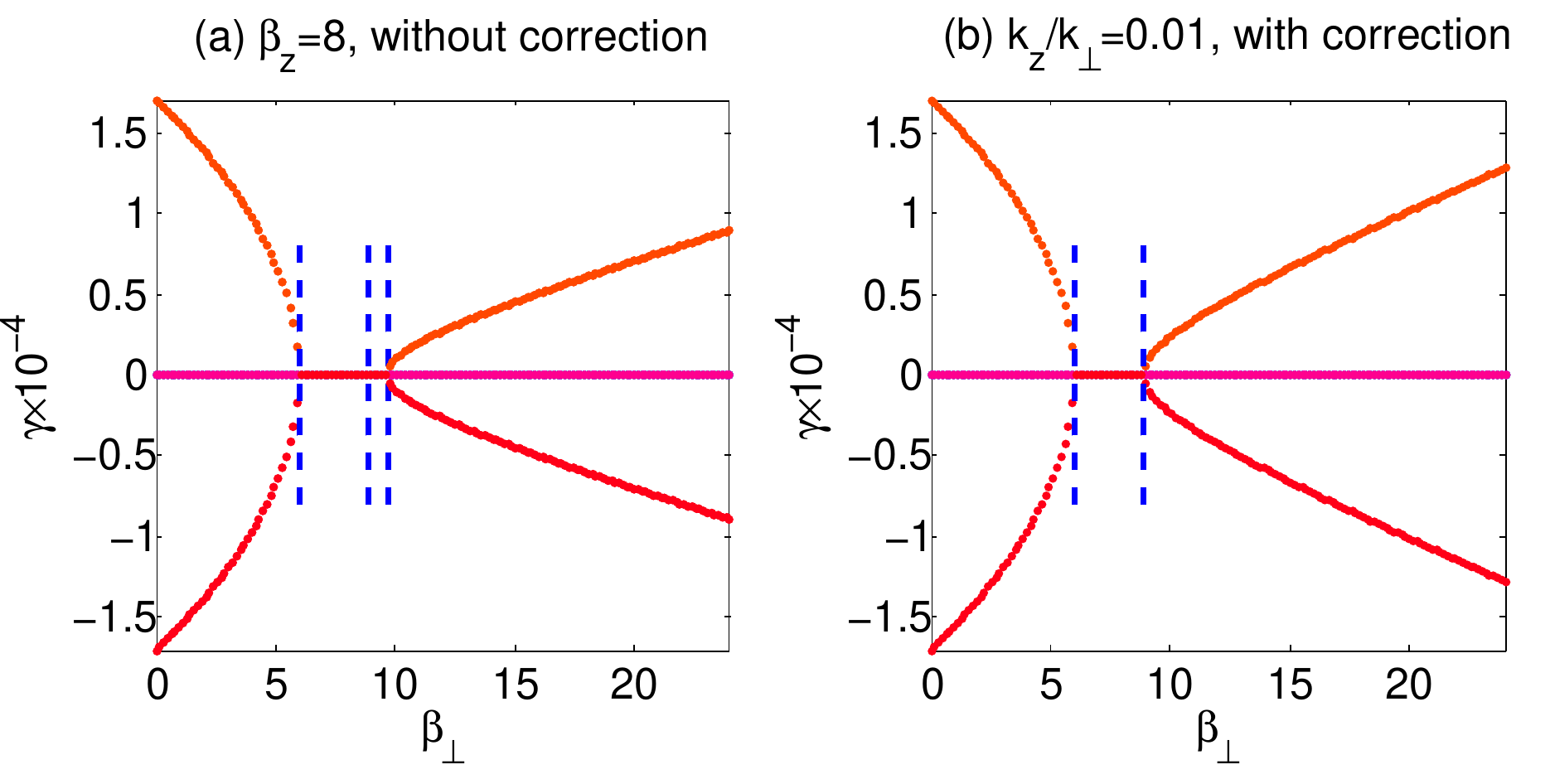}\\
  \caption{The firehose and mirror instability thresholds. The dash lines are from the analytical predictions.}\label{fig:fhmm}
\end{center}
\end{figure}

Figure \ref{fig:fhmm} shows the thresholds of the two instabilities
versus $\beta_{\perp}$ for $\beta_{\parallel,\perp e}=0$ and
$\beta_{\parallel i}=8$. The non-relativistic results for both the
with and without the factor $2$ corrected mirror instabilities are
given. For the without-correction case, the mirror instability
threshold should be $1+\sum_j\frac{\beta_{\perp
j}}{2}\big(1-\frac{\beta_{\perp j}}{\beta_{\parallel j}}\big)<0$,
which is confirmed in Fig.\ \ref{fig:fhmm}(a). We see that in
general the numerical and analytical results agree. It should be
pointed out that the CGL anisotropic fluid model (not considered)
can involve other terms originating from the transformation of
$\delta\bm B$ to $\bm \delta P$.

\subsection{Relativistic beam modes}
We now test the results of Bret \cite{Bret2007}. The ions are immobile. For the beam electrons, we have $\gamma_b=4.0$ and $n_b=0.1n_p$, where the subscripts $b$ and $p$ denote the beam and background electron quantities. For the background electrons we have
$v_p=-v_bn_b/n_p$. For $(Z_x,0,Z_z)=\bm k
v_b/\omega_{pp}=(0.3,0,3.0)$, the PDRF result $\omega^M$ and
Bret's result $\omega^B$ are listed in Table
\ref{tab:beambret}. We see that they are indeed the same for both magnetized
($B_0\neq0$, $\omega_{ce}=\omega_{pp}$) and unmagnetized ($B_0=0$)
plasmas.

\begin{table}
\begin{center}
\caption{\label{tab:beambret} Comparison of the relativistic beam
plasma solutions from the PDRF and Bret's code\cite{Bret2007}.}
\begin{tabular}{ccccccccc} \\\hline\hline
  & $\omega^M$ & 3.2736 & 3.2731 & 0.9934 & 0.3000 &  0.3000 & 0.2890-i0.1664 & 0.2890+i0.1664 \\
   $B_0=0$ & $\omega^B$ & 3.2736 & 3.2731 & 0.9934 & 0.3000 &  0.3000 & 0.2890-i0.1664 & 0.2890+i0.1664 \\
 \cline{2-9} Un-magnetized & $\omega^M$ & E-16 & 0.0000 & -0.0300 & -0.0300 &  -1.0313 & -3.2732 & -3.2736 \\
  & $\omega^B$ & 0 & 0 & -0.0300 &  -0.0300 &  -1.0313 & -3.2732 & -3.2736 \\\hline\hline
  & $\omega^M$ & 3.2945 & 3.2693 & 1.3427 & 0.5168 & 0.0771 &  0.2999+i0.0034 & 0.2999-i0.0034 \\
   $B_0\neq0$ & $\omega^B$ & 3.2945 & 3.2693 & 1.3427 & 0.5168 & 0.0771 &  0.2999+i0.0034 & 0.2999-i0.0034 \\
 \cline{2-9} Magnetized & $\omega^M$ & 0.0440 & E-16 & E-16 & -0.1019 &  -1.3983 & -3.2732 & -3.2910 \\
  & $\omega^B$ & 0.0440 & 0 & 0 & -0.1019 &  -1.3983 & -3.2732 & -3.2910 \\\hline\hline
\end{tabular}
\end{center}
\end{table}

\begin{figure}
\begin{center}
  \includegraphics[width=12cm]{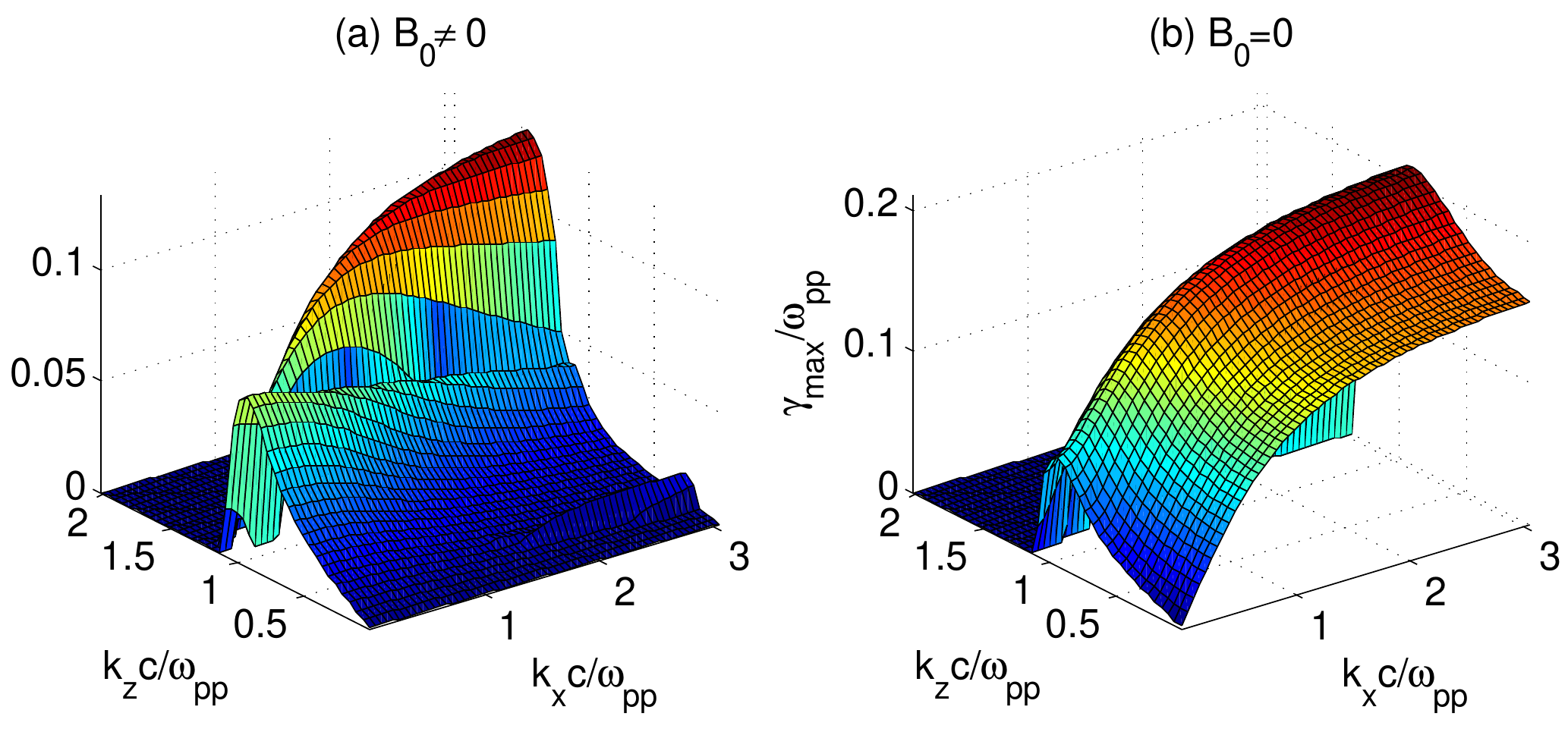}\\
  \caption{The maximum growth rate $\gamma_{\max}$ vs. $(k_x,k_z)$ for the relativistic electron beam mode with and without background $B_0$. }\label{fig:beambret}
\end{center}
\end{figure}

Figure \ref{fig:beambret} shows the maximum growth rate of the beam mode. To produce the results in Fig.\ref{fig:beambret}, the Bret method using {\it Mathematica} takes about 1 minute, whereas PDRF using {\it MATLAB} takes a few seconds.

\subsection{The Doppler effects}
In the traditional non-relativistic multi-fluid
or kinetic plasma systems, the Maxwell equations are Lorentz invariant
and the conservation equations are Galilean
invariant, so that the overall system is neither Galilean invariant
\begin{equation}\label{eq:galilean}
    \bm k'=\bm k,~~~ \omega'=\omega-\bm k\cdot \bm v_0,
\end{equation}
nor Lorentz invariant
\begin{equation}\label{eq:lorentz}
    \bm k'=\gamma(\bm k_{\parallel}-\bm v_0\omega/c^2)+\bm
k_{\perp}, ~~~ \omega'=\gamma(\omega-\bm k\cdot \bm v_0),
\end{equation}
which can lead to inaccurate treatment of the Doppler effects. For
instance, the solutions (both the frequency and the growth rate) for
the electron beam system ($v_{e0}=v_d$, $v_{i0}=0$) and the reverse ion beam
system ($v_{e0}=0$, $v_{i0}=-v_d$) would be very different, although
these two systems are equivalent under a coordinate transformation.
On the other hand, the relativistic multi-fluid plasma equations
should be Lorentz invariant.

We now consider the Lorentz invariant (\ref{eq:lorentz}) for the
single-ion ($s=e,i$) relativistic cold plasma modes. The
relativistic background density is $n_0=\gamma(v_0)n_0^0$, where
$v_0$ is the velocity of the moving frame. The initial parameters
are $m_i=4$, $m_e=1$, $q_i=-q_e=1$, $n_i=n_e=4.0$, $v_0=0$,
$v_{i0z}=v_0$, $v_{e0z}=0$, $k_xc=0.5$, and $k_zc=1.0$, which lead
to several solutions. For convenience, we consider the solution
$\omega=2.792204183976196$ (which corresponds to the largest real
$\omega$) and go to the ion moving frame with $v'_{i0z}=0$ (but
$v'_{e0z}=-v_0$), which leads to $k'_xc=0.5$ and
$k'_zc=-3.471022809144551$. The electron and ion densities then
become $n'_{e0}=9.176629354822472$ and $n'_{i0}=1.743559577416269$,
respectively. Using these new parameters, PDRF gives the largest
real part solution $\omega'_M=4.341014114998466+0i$, and
(\ref{eq:lorentz}) yields $\omega'_L=4.341014114998465+0i$.
Therefore, the difference between $\omega'_M$ and $\omega'_L$ is
only in the last decimal place.

For the $\Im\omega\neq0$ modes, the Doppler shifted wave vector is not
limited to a real number but can be complex, which is also
supported by PDRF. In principle, one can also consider the effects of nonparallel
beams, as well as that of $\nu_{ij}$ and $\bm P$, on the Doppler shift.

\subsection{Low hybrid wave polarization and gradient drift instability}
\begin{figure}
\begin{center}
  \includegraphics[width=12cm]{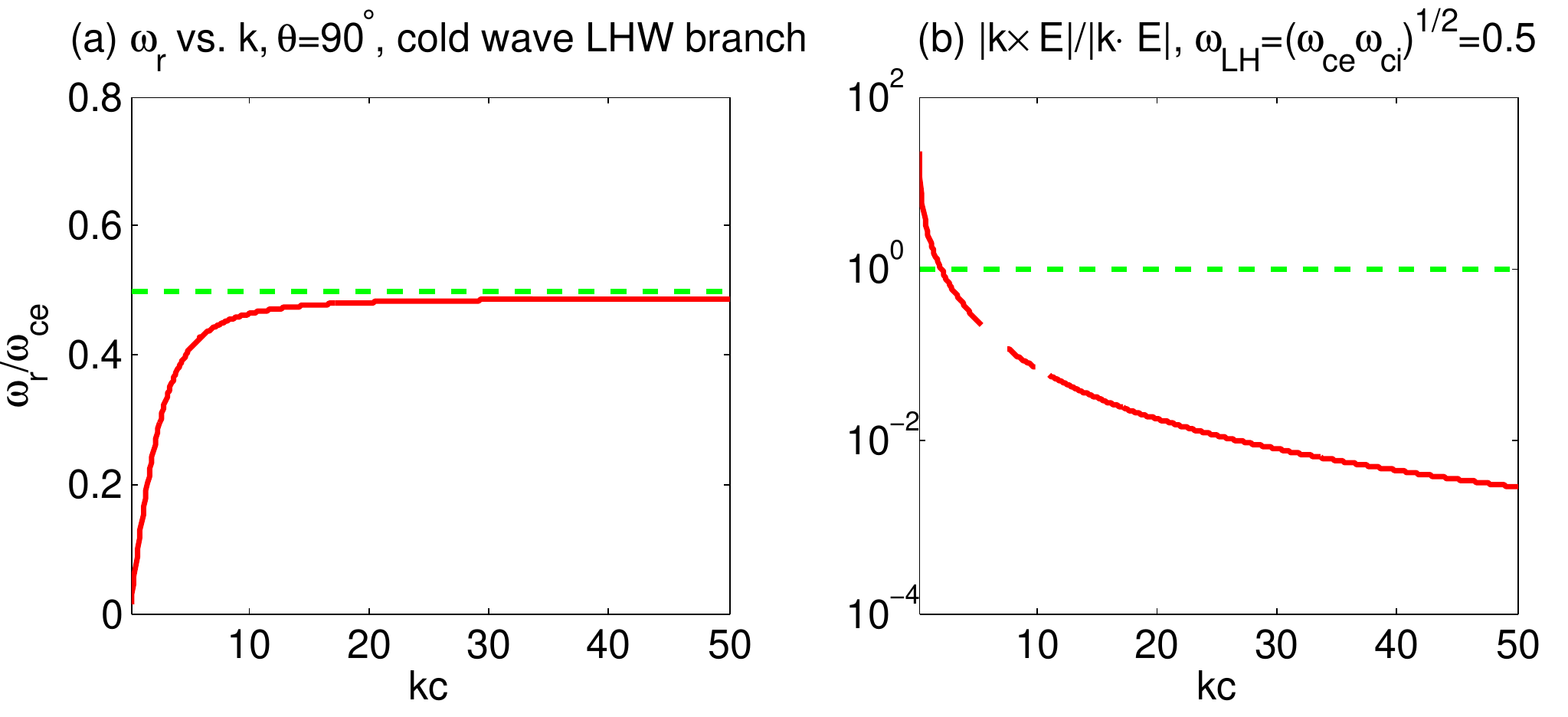}\\
  \caption{Dispersion and polarization of the quasi-electrostatic cold-plasma LHW.
  One can see that for large $kc$, (a) the numerical solution $\omega_r(k)$ (red curve) approaches the analytical value $\omega_{LH}=\sqrt{\omega_{ci}\omega_{ce}}$ (green dashed line), and (b) the expression $|\bm k\times \bm E_1|/|\bm k\cdot \bm E_1|$ (red curve) for $\omega_{LH}=0.5$ becomes very small, i.e., $\simeq 10^{-2}\ll 1$(green dashed line). }\label{fig:lhwes}
\end{center}
\end{figure}

The lower hybrid wave (LHW, $k_{\parallel}\simeq0$) is a quasi-electrostatic
mode, i.e., $|\bm k\times \bm E_1|\ll|\bm k \cdot \bm E_1|$. The
polarization property of the LHW obtained from PDRF is shown in
Fig.\ \ref{fig:lhwes}, which confirms the quasi-electrostatic
nature of the wave.

Another interesting benchmark is the LH drift wave
\cite{Krall1971,Yatsui1977},
\begin{equation}\label{eq:lhdi}
1-\frac{\omega_{pi}^2}{(\omega-kv_0)^2}+\frac{\omega_{pe}^2}{\omega_{ce}^2}\Big(1+\frac{\omega_{pe}^2}{k^2c^2}-\frac{\omega_{ce}\epsilon_n}{k\omega}\Big)=0,
\end{equation}
where $\epsilon_n$ is the gradient parameter and $v_0=v_{i0}-v_{e0}$
is the drift velocity in the electron frame, which can be due to the
density or pressure gradient or $\bm E_0\times\bm B_0$. The equilibrium drift velocity is $\bm v_{j0}=\bm v_{Dj}=-\frac{T_j}{q_jB_0}\nabla(\ln n_{j0})\times\hat{\bm b}_0$.

\begin{figure}
\begin{center}
  \includegraphics[width=16cm]{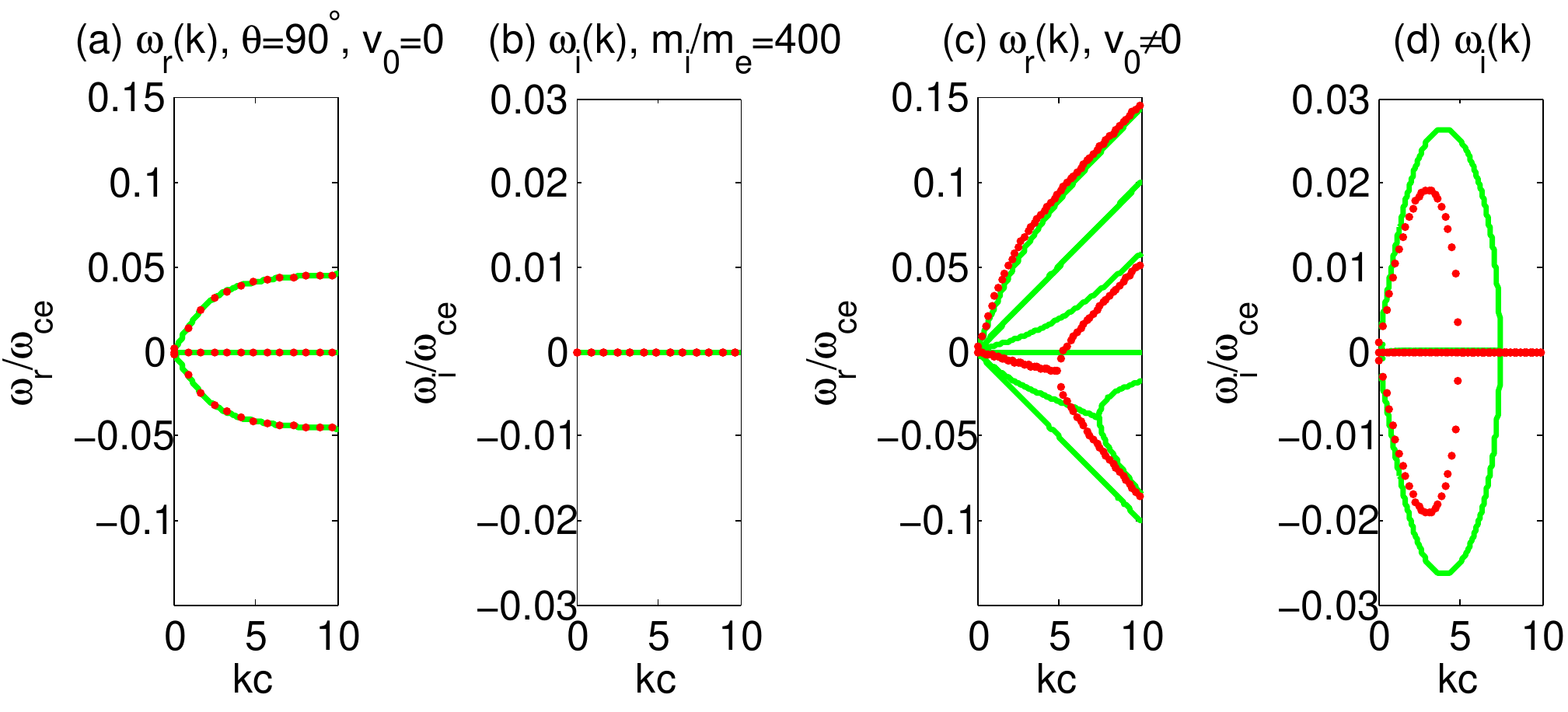}\\
  \caption{Low hybrid drift instability. Red dot lines are solutions from (\ref{eq:lhdi}) with transformation $\omega'=\omega-kv_0/2$.}\label{fig:lhdi}
\end{center}
\end{figure}

For $v_0=0$ and $\epsilon_{n}=0$, (\ref{eq:lhdi}) yields the usual
LHW solution, which is very close to the PDRF solution, as shown in
Figs.\ \ref{fig:lhdi} (a) and (b). For $v_0\neq0$, a typical
result is shown in Figs.\ \ref{fig:lhdi} (c) and (d), for $\epsilon_{niy}=\epsilon_{ney}=\epsilon_{n}=-0.15$ and
$v_{i0x}=-v_{e0x}=v_0/2=0.01$. We can see that the PDRF solutions also qualitatively
agree with (\ref{eq:lhdi}). The quantitative disagreement can be attributed to the fact that the dispersion relation (\ref{eq:lhdi}) is approximate and is derived using a Galilean transformation, whereas PDRF is more exact and is not
Galilean invariant. For example, for cold plasmas with $v_0=0$ but
$\epsilon_{n}\neq0$, the first row of the matrix $\bm M$ is zero, so that $\epsilon_{n}$ does not affect the solutions,
whereas $\epsilon_{n}$ will affect (\ref{eq:lhdi}) when $v_0=0$.

\subsection{Other effects}
In PDRF, the effects of the pressure $\bm P$, the density gradient
$\nabla n_0/n_0$ (for local inhomogeneity) and collisions $\nu_{ij}$ are given by simple models and the relativistic effects of these parameters have not been included. However, it is easy to modify the equations (\ref{eq:fpeq}) and (\ref{eq:fpeq2}), or the corresponding matrices $\bm A$ and $\bm M$, when improvement of the model equations is needed.

\section{Discussion}\label{sec:vect}
Usually one identifies a wave through its frequency. To see its
other properties such as the polarization, one should also evaluate
its eigenvectors. For example, strictly speaking, a pure shear
Alf\'en wave should not involve pressure perturbations, and a pure
ion acoustic wave (IAW) should not involve magnetic perturbations.
In the preceding section, we have used the eigenvector to verify the
quasi-electrostatic property of LHW. Information of the eigenvectors
can be useful for identifying different waves having the same or
nearly the same frequencies, for efficiently exciting specific waves
by fixing the initial condition or the source (such as an antenna),
for diagnostics of more detailed properties of the waves, etc. This
is because the perturbations in a system are given by
$X=\sum_jc_jX_j$, where $X_j$ are the eigenvectors of the eigenmode
and $c_j$ is the corresponding coefficient. The power spectrum of a
single wave parameter (e.g., $n_1$ or $B_{x1}$) cannot accurately
determine the coefficients, since different parameter values will
give different power spectrums.

With conventional dispersion relation solvers, usually only the wave frequency is obtained. One has to obtain the corresponding eigenvectors separately. PDRF provides a way to obtain the eigenvectors together with the eigenvalues.

To summarize, in this paper a general multi-fluid dispersion relation solver is provided, which shows good performance in several benchmark cases. The solver can include effects that cannot be handled by conventional solvers. The full-matrix treatment provides a general and accurate method for obtaining the dispersion relation of waves in complex fluid plasma systems. A similar treatment for kinetic plasma systems is still under development.

\section*{Acknowledgements}
The author would like to thank M. Y. Yu for improving the
manuscript. This work is partly inspired when the author was
collaborating with Liu Chen, W. Kong, Y. Lin and X. Y. Wang in the
GeFi project. Discussions and comments from Y. Xiao and Ling Chen
are also appreciated. This work is supported by Fundamental Research
Fund for Chinese Central Universities.

\appendix
\section{Extension to unmagnetized plasmas}\label{sec:pdrfu}
Setting $\bm B_0=0$ in PDRF, we can obtain
a simple unmagnetized-plasma version solver, such as shown in
Fig.\ \ref{fig:beambret}. For an unmagnetized plasma, we
need a new treatment for the anisotropic pressure, since
there is no parallel background $\bm B_0$. Here, for the pressure
$P_{\parallel j}$ and $P_{\perp j}$, the $\parallel$ and $\perp$ are
parallel and perpendicular to $\bm v_{j}$ when $\bm v_{j0}\neq 0$.
Without loss of generality, we set $\bm k=(0,0,k_z)$.

A rotation matrix
\begin{equation}\label{eq:Rxyz}
\bm R_j=\bm R_{jy}\cdot\bm R_{jx}=\left[\begin{array}{ccc}
    \cos\phi_j & 0 & \sin\phi_j\\
    0 & 1 & 0\\
    -\sin\phi_j & 0 & \cos\phi_j
    \end{array}
 \right]\cdot\left[\begin{array}{ccc}
    1 & 0 & 0\\
    0 & \cos\theta_j & -\sin\theta_j\\
    0 & \sin\theta_j & \cos\theta_j
    \end{array}
 \right]=\left[\begin{array}{ccc}
    \frac{\sqrt{v_{jy}^2+v_{jz}^2}}{v_{j}} & \frac{-v_{jx}v_{jy}}{v_{j}\sqrt{v_{jy}^2+v_{jz}^2}} & \frac{-v_{jx}v_{jz}}{v_{j}\sqrt{v_{jy}^2+v_{jz}^2}} \\
    0 & \frac{v_{jz}}{\sqrt{v_{jy}^2+v_{jz}^2}}  & \frac{-v_{jy}}{\sqrt{v_{jy}^2+v_{jz}^2}}\\
    \frac{v_{jx}}{v_{j}} & \frac{v_{jy}}{v_{j}} & \frac{v_{jz}}{v_{j}}
    \end{array}
 \right],
\end{equation}
is required to transform $\hat
\parallel_j$ and $\hat \perp_j$ to $\hat{x}$, $\hat{y}$, $\hat{z}$, i.e., the pressure tensor should be
\begin{equation}\label{eq:Ptensor}
\bm P_j=\bm R_j^T\cdot\left[\begin{array}{ccc}
    P_{\perp j} & 0 & 0\\
    0 & P_{\perp j} & 0\\
    0 & 0 & P_{\parallel j}
    \end{array}
 \right]\cdot\bm R_j,
\end{equation}
which yields
\begin{eqnarray}\label{eq:Pjxyz}
  P_{jxx} =
  P_{\perp j}+(P_{\parallel j}-P_{\perp j})\frac{v_{jx}^2}{v_{j}^2}
  &,&
  P_{jyy} =
  P_{\perp j}+(P_{\parallel j}-P_{\perp j})\frac{v_{jy}^2}{v_{j}^2},\cr
  P_{jzz} =
  P_{\perp j}+(P_{\parallel j}-P_{\perp j})\frac{v_{jz}^2}{v_{j}^2} &,&
  P_{jxy} =
  P_{jyx}=(P_{\parallel j}-P_{\perp j})\frac{v_{jx}v_{jy}}{v_{j}^2},\cr
  P_{jzx} =
  P_{jxz}=(P_{\parallel j}-P_{\perp j})\frac{v_{jx}v_{jz}}{v_{j}^2} &,&
  P_{jzy} = P_{jyz}=(P_{\parallel j}-P_{\perp j})\frac{v_{jy}v_{jz}}{v_{j}^2}.
\end{eqnarray}
For $\bm v_{j0}=0$, one should set $\bm P_j=P_j$ as a scalar
quantity.

Treatment for other unmagnetized plasmas are similar to that for
the magnetized plasma. The above description provides a guide for
developing a more general unmagnetized plasma version of PDRF.

\section{PDRF User Manual}\label{sec:manu}
PDRF consists of two files: the main program ``pdrf.m" and the input data
file ``pdrf.in". The input file has the follow structure
\begin {verbatim}
qs      ms      ns      vsx      vsy      vsz      csz      csp     epsnjx   epsnjy
-1.0    1.0     4.0     0.0      0.0      0.0      0.0      0.0     0.0      0.0
1.0     4.0     4.0     0.0      0.0      0.0      0.0      0.0     0.0      0.0
\end {verbatim}
One can add the corresponding parameters for additional species.

Normalizations and other parameters can be reset in ``pdrf.m". Since PDRF is a
short code, one can also easily modify the default setup for other applications.

\end{document}